\documentclass[showpacs,twocolumn,amsmath,amssymb,pre,aps]{revtex4} 
\usepackage{graphicx}
\usepackage{bm}
\usepackage{mathrsfs}

\begin{document}
\title
{The dynamics of colloids in a narrow channel driven by a non-uniform force}
\draft

\author{D.~V.~Tkachenko$^{1}$, V.~R.~Misko$^{1}$, 
and F.~M.~Peeters$^{1,2}$}
\affiliation
{$^{1}$Department of Physics, University of Antwerpen, Groenenborgerlaan 171, B-2020 Antwerpen, Belgium}
\affiliation
{$^{2}$Departamento de F\'{\i}sica, Universidade Federal do 
Cear\'{a}, 60455-900 Fortaleza, Cear\'{a}, Brazil} 

\date{\today}

\begin{abstract}

Using Brownian dynamics simulations, we investigate the dynamics of colloids confined in 
two-dimensional narrow channels driven by
a non-uniform force $F_{dr}(y)$. 
We considered linear-gradient, parabolic and delta-like 
driving-force profiles. 
This driving force induces melting of the colloidal solid 
(i.e., shear-induced melting),
and the colloidal motion experiences a transition from elastic to  plastic regime with increasing $F_{dr}$. 
For intermediate $F_{dr}$ (i.e., in the transition region) the response of the system, i.e., the distribution of the velocities 
of the colloidal chains $\upsilon_i(y)$, in general does not coincide with the profile of the driving force $F_{dr}(y)$, and depends
on the magnitude of $F_{dr}$, the width of the channel and the density of colloids. 
For example,
we show that the onset of plasticity is first observed near the boundaries while the motion in the central region is elastic. 
This is explained by: 
i) (in)commensurability between the chains due to the larger density of colloids near the boundaries, and 
ii) the gradient in $F_{dr}$.
Our study provides a deeper understanding of the dynamics of colloids in channels and could be accessed in experiments on colloids 
(or in dusty plasma) with, e.g., asymmetric channels or in the 
presence of a gradient potential field. 

\end{abstract}
\maketitle

\section{Introduction} 

During the last decade there has been a growing interest in the research of physical properties of colloidal systems.
This interest is by part due to perspectives of practical use 
of colloids, e.g., in biology, medicine, meteorology, 
food production, etc.
On the other hand, colloids serve as a convenient model system for studying, e.g., phase transitions \cite{Bechinger,Han,Nielaba}, diffusion \cite{Reichhardt}, or commensurate-incommensurate transitions \cite{Achim}.
Typical dimensions of colloids are in the micrometer regime and 
their dynamics is governed by a time scale in the microsecond regime and therefore is suitable for direct observation in real space 
and time.
Furthemore, it has been possible to tune the inter-particle interaction potential which opens thereby a wide area for research 
of fundamental properties of classical systems.

In Refs. \cite{Doyle,Ricci,Stratton} structural properties of magnetic mono-colloidal mixtures confined in narrow
two-dimensional (2D) channels were studied under equilibrium conditions. 
Structural deviations from an infinite 2D crystal and related to 
it oscillations in structural properties were predicted.
Transport properties of paramagnetic mono-colloidal mixtures in narrow channels being in
non-equilibrium but under stationary and homogeneous external conditions (in the gravity field) were investigated in 
Refs. \cite{Leiderer,Leiderer2}.
It was shown that gravitation action results in the occurrence 
of a density gradient along the channel, that in its turn leads 
to a gradient in the number of chains.
The latter was predicted for a driven colloidal system in the presence of a constriction in the channel \cite{Peeters}. 
Besides, authors of Refs.~\cite{Leiderer,Leiderer2} also studied the relation between velocities of colloidal motion, diffusive behavior 
and the self-organized order in the system.

The transition from a hexagonal to a chain-like ordered phase was studied in 
Ref.~\cite{Peeters3} for Yukawa particles (applicable for charged colloids 
and dusty plasma \cite{piel,mikovi,melzer}) confined in a two-dimensional parabolic channel.
It was shown that the system crystallized in a number of chains similar to a colloidal system confined in a narrow channel. 
The authors of Ref.~\cite{Peeters3} analyzed the structural transitions between 
the ground state configurations and showed that such system exhibited a rich phase 
diagram at zero temperature under continuous and discontinuous structural transitions. 

The purpose of this study is to achieve a deeper understanding of the influence of structural
properties on the motion of paramagnetic colloids in narrow channels under non-uniform distributed driving force. 
In particular, we consider:
(i) a linear increasing driving force from one to the other boundary of the channel, i.e., a linear-gradient driving force, and
(ii) a parabolic-profile driving force, i.e., equal to zero at the boundaries and maximum at the center of the channel.
Although the experimental realization of such profiles is not as obvious as for the simplest case of a constant driving force 
(created, e.g., by gravity in an inclined channel), 
they can serve as a model of a very important case of a channel modulated in a non-uniform manner, e.g., by a non-uniform distributed 
disorder (``pinning'') or surface roughness. 
The combination of a uniform force applied to colloids with a 
non-uniform modulation in such a channel will result in an effective non-uniform driving, i.e., $F_{dr,nu} = F_{dr,u} - f_{m,nu}$.
The spatial period of the modulation should be chosen, obviously, essentially smaller than the colloidal radius. 
For example, if the modulation depth of a channel linearly grows from one edge of the channel to another, the effective driving force distribution is described by a linearly growing function.
Moreover, the application of a non-uniform driving force to a discrete system of interacting
particles confined to a quasi-1D channel is expected to result in a rich dynamics interesting
from the point of view of fundamental research, and it could be useful for understanding
the dynamics of other interacting systems moving in narrow channels.

The paper is organized as follows.
The model is described in Sec.~II.
In Sec.~III, we present the results of our numerical calculations of the response of the colloidal 
system to a non-uniform, i.e., linear-gradient and parabolic, driving force.
In Sec.~IV, we study the mobility of stripes and dynamical phases of their motion.
The conclusions are presented in Sec.~V.

\section{The model and method} 

Motion of the system of paramagnetic colloids in a 
two-dimensional (2D) narrow channel under the action of a non-uniform 
distributed driving force $F_{dr}(y)$ is investigated using 
the Langevin equations of motion in the overdamped regime, 
i.e., the Brownian dynamics (BD) method.
It is supposed that a magnetic field is applied perpendicularly to the channel plane
(i.e., perpendicular to the $xy-$plane).
Following Ref.~\cite{Leiderer}, we choose $B_0=1\ mT$.
In this case the magnetic field induces magnetic moments in the colloids $M_0=\chi B_0$
($\chi=3\cdot10^{-11}$ \cite{Leiderer}) and, therefore, the interparticle interaction 
is described by the dipolar repulsive potential $V(r_{ij})$ 
(for $r_{ij} \geq r_{col}$, and by hard core interaction for 
$r_{ij} < r_{col}$): 
%
\begin{equation}\label{L1}
 V(r_{ij})= 
\left\{ 
\begin{array}{cc}
\frac{\mu_0}{4\pi}\frac{M_0^2}{r_{ij}^3}, & \ \ \ {\rm if } \ \ \ r_{ij} \geq r_{col}, \\ 
\infty, & \ \ \ {\rm if } \ \ \ r_{ij} < r_{col}, 
\end{array}
\right.  
\end{equation}
where $\mu_0$ is a magnetic constant, 
$\vec{r}_{ij}$ is the distance between $i$th and $j$th particles.
Following Ref.~\cite{Leiderer}, we have chosen the colloidal radius 
$r_{col}=2.275\ \mu m$. 
The diameter of a colloidal particle is denoted as 
$\sigma_{col} = 2 r_{col}$. 

The use of the BD method assumes that colloids are considered in 
the limit of large viscosity. 
The system of overdamped equations of motion in this case is given 
by 
\begin{equation}\label{L2}
\eta\frac{d\textbf{r}_i}{d t}=-\sum_{j,j\neq i}{\nabla \textbf{V}}(\textbf{r}_{ij})+\textbf{F}_{dr}(y_i)
+\textbf{F}_{i}^{T},
\end{equation}
where the friction constant $\eta = 1.2275 \cdot 10^{-8}$ is Stokes friction for a spherical particle of radius $r_{col}$ 
\cite{Leiderer}, 
$\textbf{r}_i$ is the position vector of i$th$ colloid. 
An external non-uniform stationary force $\textbf{F}_{dr}(y_i)$ 
is applied along the channel 
and depends only on the transverse coordinate $y$. 
The thermal stochastic term $\textbf{F}_{i}^{T}$ entering 
Eq.~(\ref{L2}) obeys the following conditions: 
\begin{equation}
\langle F_i^T(t)\rangle =0
\end{equation}
and
\begin{equation}
\langle F_i^T(t) F_j^T(t^\prime)\rangle =2\eta k_B T\delta_{ij}\delta(t-t^\prime).
\end{equation}

To find the ground state (or an initial state with the lowest free energy close to the ground state) 
of the system, we first solve Eq.~(\ref{L2}), for nonzero temperature in the absence of driving, 
thus simulating annealing process used for obtaining the ground state of, e.g., colloids 
\cite{annealing} 
or vortices in a superconductor (see, e.g., 
\cite{prlpenrose,prldisks,prlcorbino}). 
Then we set temperature equal to zero and solve Eq.~(\ref{L2}) for colloids driven by the external force $F_{dr}(y_{i})$. 
Note that inclusion of temperature fluctuation would lead to the degradation of the 
colloidal stripe structure and, as a result, to smearing of the velocity profile of different colloidal chains. 
We focus on the features of the velocity profiles related to the structure of the chains 
driven by a non-uniform force, and we neglected temperature-induced 
fluctuations in our calculations. 
In our numerical simulations, we use a simulation cell with sizes equal to the channel width in the $y$-direction and typically 
$140r_{col}$ along the channel, i.e., in the $x$-direction. 
We use periodic boundary condition along the $x$-direction.
Interaction of colloids with lateral walls is hard-wall. 
The system of overdamped equations (\ref{L2}) is solved using a variable time step,
calculated from the condition $max{\{d\textbf{r}_i\}}\leq \delta$, where
$\textbf{dr}_i$ is the displacement of $i$th colloid, and $\delta$ was typically chosen
$r_{col}/200$ or $r_{col}/100$, although in most cases the value 
$r_{col}/20$ was sufficient.

\section{Response of the system to non-uniform driving force} 

\subsection{The ground state configurations} 

As was shown in Refs.~\cite{Doyle,Peeters,Stratton}, in the absence of external forces interacting colloids in narrow 2D channels are 
ordered in stripe structures.
Therefore, one might expect that appliyng an external force along 
the channel will result in an ordered drift of colloidal stripes 
in the channel.
(This obviously applies to the case of rather weak driving force or its gradient and very narrow channels.
Strong inhomogeneous driving in a wide enough channel could lead to the appearance of instabilities
in the transverse direction to the applied driving and thus to the migration of colloids between stripes.) 
Thus the investigation of the motion of colloids in narrow channels can be reduced to the analysis of the drift of colloidal stripes.
Note that in this case the order of individual colloids in a stripe does not change with time, and the motion of colloids in stripes resembles single-file diffusion 
(see, e.g., \cite{Wei,Kollmann,Fabioprl,Coupier,eplsfd}).

\begin{figure}[btp]
\begin{center}
\hspace*{-0.5cm}
\includegraphics*[width=8.5cm]{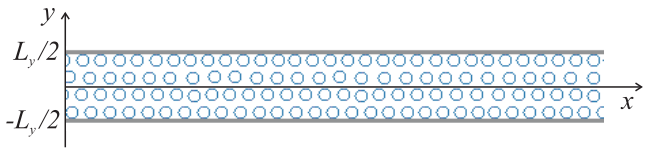}
\end{center}
\vspace{-0.5cm}
\caption{
(Color online)
A typical view of a highly-ordered ground-state ($\Psi_{n_l} = 0.96$) colloidal distribution
in a channel, for the total density $\rho = 0.6 \sigma_{col}^{-2}$.
}
\label{FigGS}
\end{figure}

For studying the colloidal motion in channels we choose such ground state configurations which are characterized by strongly pronounced stripe structure and thus by a high value of stripe order
parameter \cite{Leiderer}
\begin{equation}\label{L3}
 \Psi_{n_l}=\frac{1}{N}\left|\sum_{j=1}^N e^{i2\pi(n_l-1)y_i/L_y}\right|.
\end{equation}
The order parameter $\Psi_{n_l}=1$ for particles distributed equidistantly in $n_l$ chains across the channel, 
and $\Psi_{n_l} \ll 1$ for a non-chained case.
Here $L_y$ is the channel width, $y_i$ is the $y$ coordinate of $i$th colloid and $N$ is the number of particles in the system. 
Such a choice of the initial configurations predetermined 
a well-defined stripe structure and allowed
us to focus on the study of stripe dynamics rather than on the dynamics of a low-ordered phase
(i.e., for $\Psi_{n_l} \ll 1$).
A typical configuration of the initial colloidal distribution is shown in Fig.~\ref{FigGS}.
Our estimate of the inter-chain interaction shows that the main contribution is given only
by the adjacent stripes while the contribution from more remote stripes is negligible.

The profile of the applied driving force and the structural distribution of colloids in a channel
are the two most important factors defining the dynamics of colloidal motion in a narrow two-dimensional channel in the overdamped regime.
Their roles essentially distinguish depending on the value of the applied driving.
For an overall strong drive the distribution of the driving force 
$F_{dr}(y)$ is dominating.
Contrarily, in case of weak drive, the motion of colloidal stripes 
is mainly determined by the structural distribution of colloids.
In intermediate cases the character of colloidal motion results from a complex competition between these two factors. 

The distribution of the density of colloids in the transverse direction $\rho_{chain}(y)$ is essentially non-uniform, i.e., 
presence of the channel boundaries leads to a relatively higher 
colloidal density on the periphery and lower in the central chains 
\cite{Doyle,Stratton,Leiderer}.
The distribution of colloidal density becomes more homogeneous with increasing channel width $L_y$, and in the limit of a wide channel, 
$L_y \gg \sigma_{col}$, the colloidal structure corresponds 
to a regular hexagonal lattice. 

Another essential feature of the colloidal distribution in a channel consists in oscillating dependence
of the defect concentration on the channel width $L_y$ \cite{Doyle}.
Thus increase of the channel width is periodically accompanied by a qualitative rearrangement of the
colloidal structure, namely, by the occurrence of new chains.
The values of the channel width, 
for which the colloidal structures rearrange, 
correspond to low-ordered states ($\Psi_{n_l} \ll 1$) 
when the stripe structure is hardly distinguishable.
This fact is reflected in the oscillation of the stripe order parameter $\Psi_{n_l}$ with increasing width of the channel. 
It is worth to note however that the fact that $\Psi_{n_l}$ is large does not mean that the defect concentration is low. 
Actually it means only that stripes are well pronounced and all defects are mainly topological.

\subsection{The integral of motion} 

Before discussing results of our numerical calculations,
we note a useful property that directly follows from system 
of equations (\ref{L2}), namely, the integral of motion: 
\begin{equation}\label{L4}
 \eta\sum_i\frac{d\vec{r}_i}{d t}=\sum_i\vec{F}_{dr}(y_i).
\end{equation}
Separating the longitudinal component we obtain, 
in terms of average stripe velocities,
\begin{equation}\label{L5}
 \eta\sum_n\bar{\upsilon}_{xn}=\sum_n F_{dr}(y_n),
\end{equation}
where $n$ denotes the $n$th chain, 
$\bar{\upsilon}_{xn}$ is the average velocity of $n$th stripe. 
This expression has a clear meaning: the sum of all average chain 
velocities is defined by the momentum of the external force 
transferred to the system. 
From Eq.~(\ref{L5}) it follows that the colloidal system is in
the static state, i.e., does not move, only in case of: 
(i) zero driving force;
(ii) sign-alternating driving force satisfying the condition
$\sum_n F_{dr}(y_n)=0$ 
and $max \mid F_{dr}(y_n) \mid < F_{c}$, where $F_{c}$ is some critical force. 

This result is remarkable due to the fact that it differs from the
case of particles moving under the action of an external force in
any stationary periodic potential.
In the latter case, for a commensurate configuration (i.e., when
the number of colloids per unit length coincides with the number
of potential minima), there is always a threshold value of the
force related to the ``static friction''.
Thus the system of particles moves if the driving force exceeds 
the threshold value. 
In our case, the system of colloids moves always when the sum of 
forces in (\ref{L5}) is larger than zero. 

Note that the conservation law (\ref{L5}) is fulfilled for any 
distribution of the driving force $F_{dr}(y)$ and it is not valid 
if temperature is non zero.

\subsection{Stripe velocity versus non-uniform driving force} 

In this subsection, we study a response of the colloidal system confined in a channel to an external non-uniform driving force.
Clearly, this case is more general than the earlier studied
case of a uniform driving created, e.g., by the gravity in 
inclined channels \cite{Leiderer}. 
A gradient driving force produces a shear stress that allows us to
examine elastic properties of the colloid ``lattice'' and to reveal
the onset of plastic motion.
The transition from elastic to plastic motion leads to a number of
different dynamical regimes.

Without loss of generality, we consider two typical profiles of the  non-uniform distributed driving force:
(i) a linear distributed force given by
\begin{equation}\label{L6}
 F_{dr}(y)=F_{max}\left(\frac{1}{2}+\frac{y}{L_y}\right),
\end{equation}
and (ii) a parabolic distributed driving
\begin{equation}\label{L7}
 F_{dr}(y)=F_{max}\left(1-\left(\frac{2y}{L_y}\right)^2\right),
\end{equation}
where $F_{max}=mg\rm{sin}(\alpha)$ is the maximum value of the driving force which we express, for convenience, via the projection 
of the gravity force that acts on a free colloidal particle being 
on an inclined plane at angle $\alpha$; 
$m$ is the mass of the colloidal particle for the density of colloids 
$\rho_{col}=1600\ kg/m^3$ \cite{Leiderer}.

\begin{figure}[btp]
\begin{center}
\vspace*{0.5cm}
\includegraphics*[width=8.5cm]{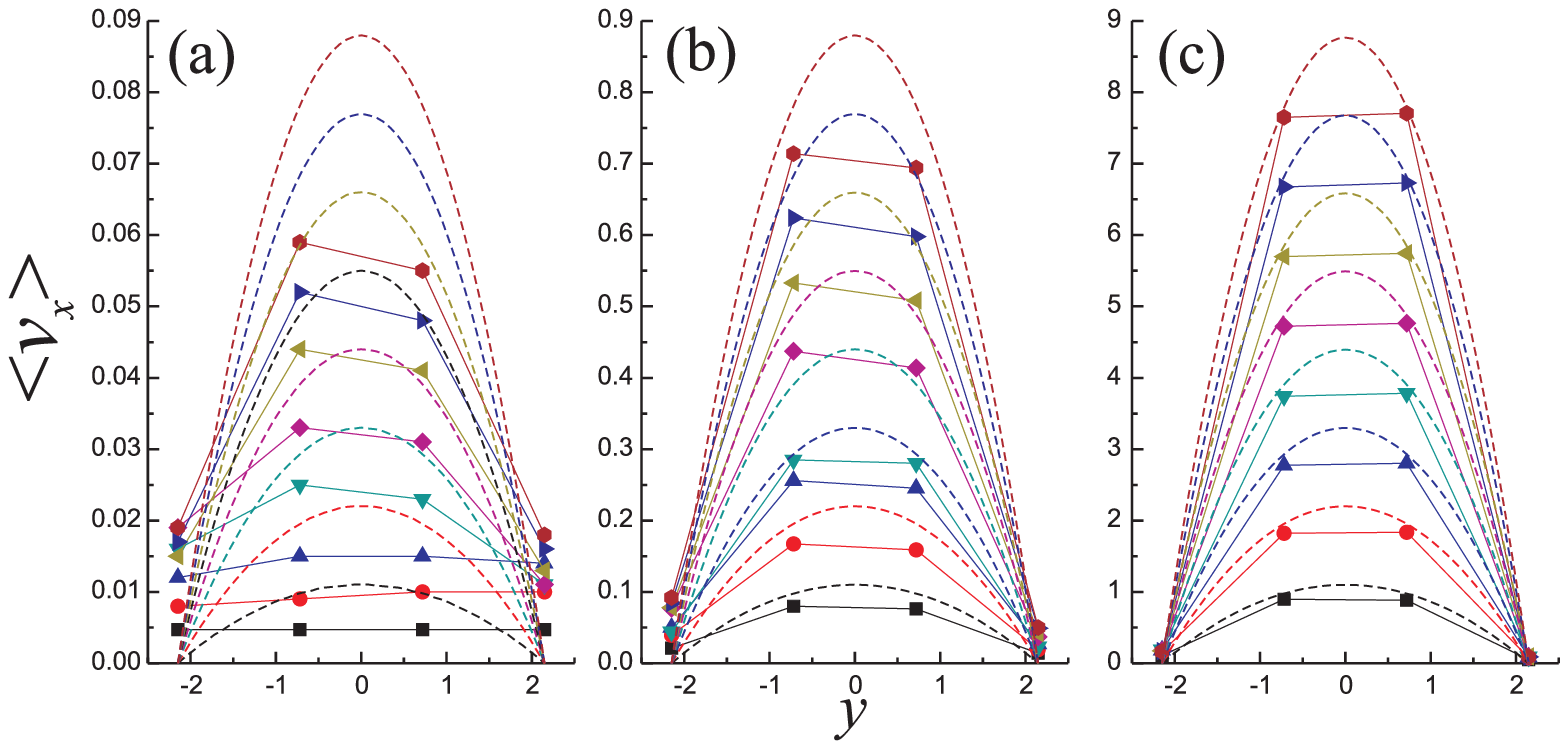}
\includegraphics*[width=8.5cm]{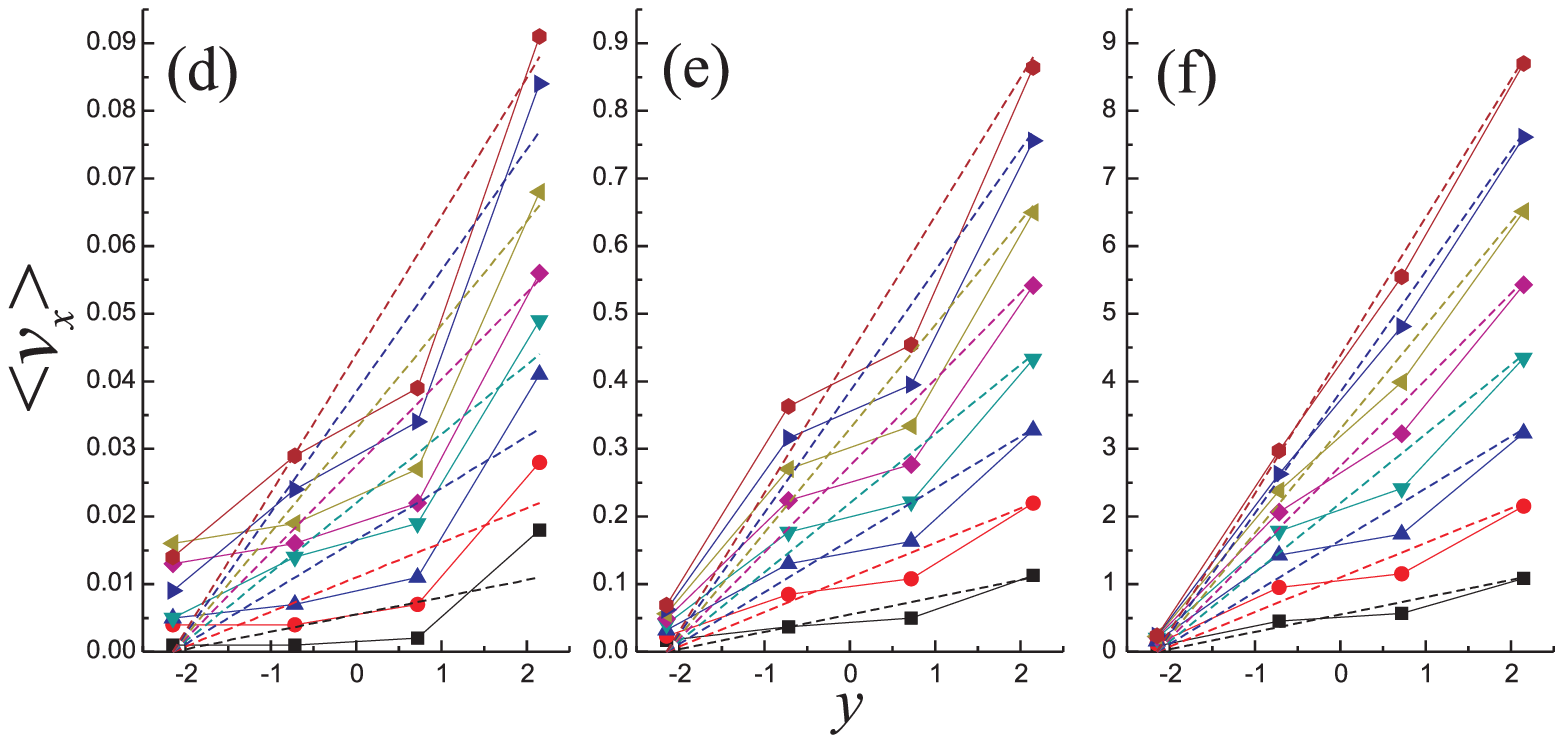}
\end{center}
\vspace{-0.5cm}
\caption{
(Color online)
The profile of average chain velocities $\langle\upsilon_x\rangle$
as a function of
the transverse coordinate $y$ for a parabolic ((a) to (c)) and
a linear ((d) to (f)) profiles of the applied driving force
for various values of $\alpha$:
$\alpha = 0.01$ to 0.08 ((a), (d)),
$\alpha = 0.1$ to 0.8 ((b), (e)), and
$\alpha = 1$ to 8 ((c), (f)).
The width of the channel is $L_y/\sigma_{col} = 4.3$ and the total
colloidal density is $\rho=0.6\ \sigma_{col}^{-2}$.
The symbols connected by solid lines are average velocities
of chains while dashed lines show velocities of (non-interacting)
chains corresponding to the distribution of the applied force.
}
\label{FigLinParVy}
\end{figure}

The average velocity of different chains $\bar{\upsilon}_{x}$ as a
function of their transverse coordinate $y_n$ is presented
in Fig.~\ref{FigLinParVy} for values of $\alpha$ varying in the range
$0.01\leq\alpha\leq8$.
The profiles of the calculated average stripe velocity
$\bar{\upsilon}_{x}(y_n)$ are shown by different symbols (corresponding to different values of driving force expressed 
in terms of angle $\alpha$, -- see the figure captions) 
connected by solid lines. 
For comparison, the profiles of stripe velocities in the absence of
interaction between the stripes (i.e., renormalized applied external
force $F_{dr}(y)/\eta$) are shown by dashed lines, for different magnitudes of the driving force. 
With increasing $\alpha$ and, accordingly, the total amplitude 
$F_{max}$
of the driving force the role of the shape of the distribution of
applied force on $\bar{\upsilon}_{x}(y_n)$ increases, while the influence
of the structure of colloids and the density of individual chains,
$\rho_{chain}(y)$, on the contrary, decreases.
In the limit of very large driving force $F_{dr}$ 
($1\lesssim\alpha\leq8$),
the distribution of stripe velocity is completely defined by the profile of
the applied force $F_{dr}(y)$, as one can expect in the plastic limit
(see Figs.~\ref{FigLinParVy}(c),(f), for parabolic and 
linear-gradient driving, correspondingly). 
On the contrary, for small amplitudes $F_{max}$ of the driving force
($0.01\leq\alpha\lesssim0.05$), deviations of $\bar{\upsilon}_{xn}$ from the profile of the driving force $F_{dr}(y)$ are essential
(Fig.~\ref{FigLinParVy}(a),(d)).
This limit corresponds to the elastic or quasi-elastic regime.
In the intermediate region of $F_{dr}(y)$, $0.05\lesssim\alpha\lesssim1$,
the function $\bar{\upsilon}_{x}(y_n)$ is essentially influenced by
both factors, $F_{dr}(y)$ and $\rho_{chain}(y)$.

The above behavior is rather general and is typical, in principle,
for any profile of the external driving. 
At the same time, the shape of the driving-force profile is important, especially, in the region of intermediate driving forces. 
In other words, for the same set of parameters (i.e., the total density, $\rho$, the channel width, $L_y$, and $\alpha$) the difference between $F_{dr}$ and $\bar{\upsilon}_{x}$ varies for 
parabolic and linear profiles of the external driving force. 

In order to introduce a measure of conformity of the profiles of the
calculated average velocity $\bar{\upsilon}_{x}(y)$ and the external
force $F_{dr}$, we define a normalized standard deviation $\Delta$ 
in the form:
\begin{equation}\label{L8}
\Delta=\sqrt{\sum_{chain}\left(\frac{F_{dr}(y_{chain})/\eta -
v_{chain}}{F_{max}}\right)^2}.
\end{equation}
Low values of $\Delta$ correspond to small difference between
$\bar\upsilon_{xn}(y)$ and $F_{dr}(y)$.
For a parabolic driving-force profile, the gradient of the applied force and thus the shear stress between inner chains is much less 
than between those at the periphery, while for a linear profile the 
gradient is constant.

\begin{figure}[btp]
\begin{center}
\vspace*{0.5cm}
\hspace*{-0.5cm}
\includegraphics*[width=5.5cm]{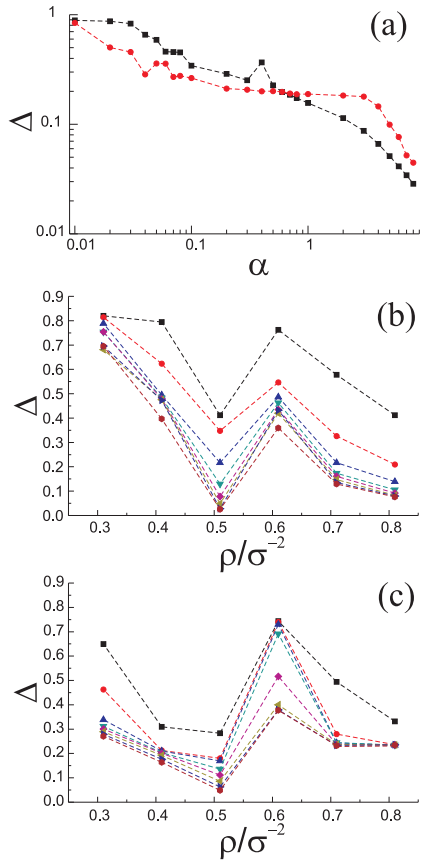} 
\end{center}
\vspace{-0.5cm}
\caption{
(Color online)
The normalized standard deviation $\Delta$ as a function of $\alpha$, 
calculated from data presented in Fig.~\ref{FigLinParVy} 
(i.e., for a four-chain structure calculated for 
$\rho=0.6\ \sigma_{col}^{-2}$). 
Squares (connected by a dot-dashed line) correspond to a linear 
profile of the driving force; circles correspond to a parabolic 
profile (a). 
The function $\Delta$ vs. the total colloidal density $\rho$ for: 
a parabolic driving force (b) and 
a linear-gradient driving force (c), 
for different values of $\alpha = 0.1$ to 0.8 
(i.e., for three- and four-chain structures as shown in 
Figs.~4 and 5). 
Here we use the same symbols as in Fig.~\ref{FigLinParVy} to denote 
velocities of different colloidal stripes. 
}
\label{FigStD}
\end{figure}

The results of our calculation of the standard deviation $\Delta$
are shown in Fig.~\ref{FigStD} versus driving force (a) and 
density (b,c), for linear and parabolic profiles of the driving 
force.

Note that for small driving forces, i.e., for elastic motion,
the deviation of the velocity profile $\bar\upsilon_{xn}(y)$ from
that of the applied driving $F_{dr}(y)$ is less in case of a 
linear driving-force profile. 
This means that the elastic deformation of the colloidal ``solid''
related to a linear driving is stronger than that for a parabolic 
driving. 
Contrary, for overall large driving (i.e., $\alpha$), 
linear-gradient-driven stripes follow the profile of the driving
force to a lesser extent than in case of parabolic driving
(see Fig.~\ref{FigStD}(a)).
This means that the average dynamical friction between chains 
is larger in case of a linear-driving applied force. 
This conclusion seems to be quite surprising keeping in mind the large gradient of the driving force (experienced by peripheral 
chains) in case of a parabolic driving. 
As we show below, it can be understood in terms of incommensurability 
between peripheral chains (i.e., there the driving gradient is maximum) 
related to different densities of colloids in different chains. 
The function $\Delta(\rho)$ versus the overall colloidal density is presented
in Figs.~\ref{FigStD}(b) and~\ref{FigStD}(c), respectively, for parabolic and
linear-gradient driving, for different $\alpha$, and 
$L_{y}/\sigma_{col} = 4.3$.
Note that in both cases $\Delta(\rho)$ has a minimum at about 
$\rho = 0.5 \sigma_{col}^{-2}$ 
corresponding to a perfect chained structure and incommensurate number of colloids in different chains. 
As a result, the chains easily slide with respect to each other, 
and the velocity profile reproduces that of the driving force 
(if the driving is strong enough, see Figs.~\ref{FigStD}(b) and (c)). 
The minimum in $\Delta(\rho)$ at $\rho = 0.5 \sigma_{col}^{-2}$ 
is followed by a sharp increase of $\Delta(\rho)$ at about 
$\rho = 0.6 \sigma_{col}^{-2}$ explained by the appearance of 
defects in the chained structure. 
The defects lock the motion of different chains, similarly to the 
case of low densities (i.e., $\rho \sim 0.3 \sigma_{col}^{-2}$), 
and the system displays a rigid-body(RB)-like motion. 
With increasing $\rho$, the maximum in $\Delta$ alternates with 
a gradual decrease indicating increasing ordering of colloids 
in stripes. 
Note that the behavior of $\Delta$ is in general similar for the 
linear and parabolic driving force (i.e., for high densities $\rho$), 
the absolute values and dispersion of $\Delta$ for different 
$\alpha$ 
are lower in case of the linear-gradient driving, in agreement
with the result shown in Fig.~\ref{FigStD}(a).

\subsection{The role of colloidal density in stripes} 

\begin{figure}[btp]
\begin{center}
\vspace*{0.5cm}
\includegraphics*[width=8.5cm]{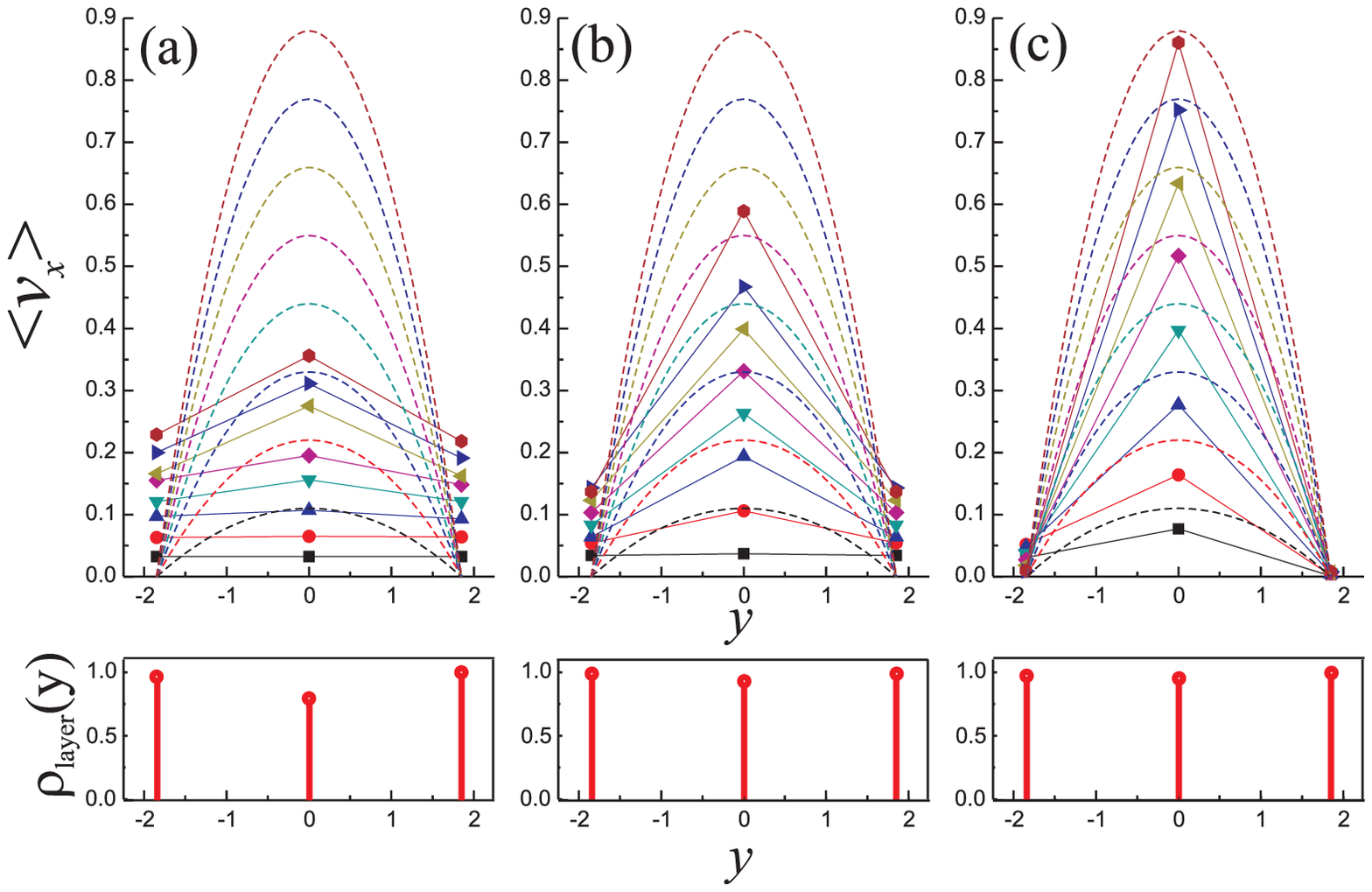}
\includegraphics*[width=8.5cm]{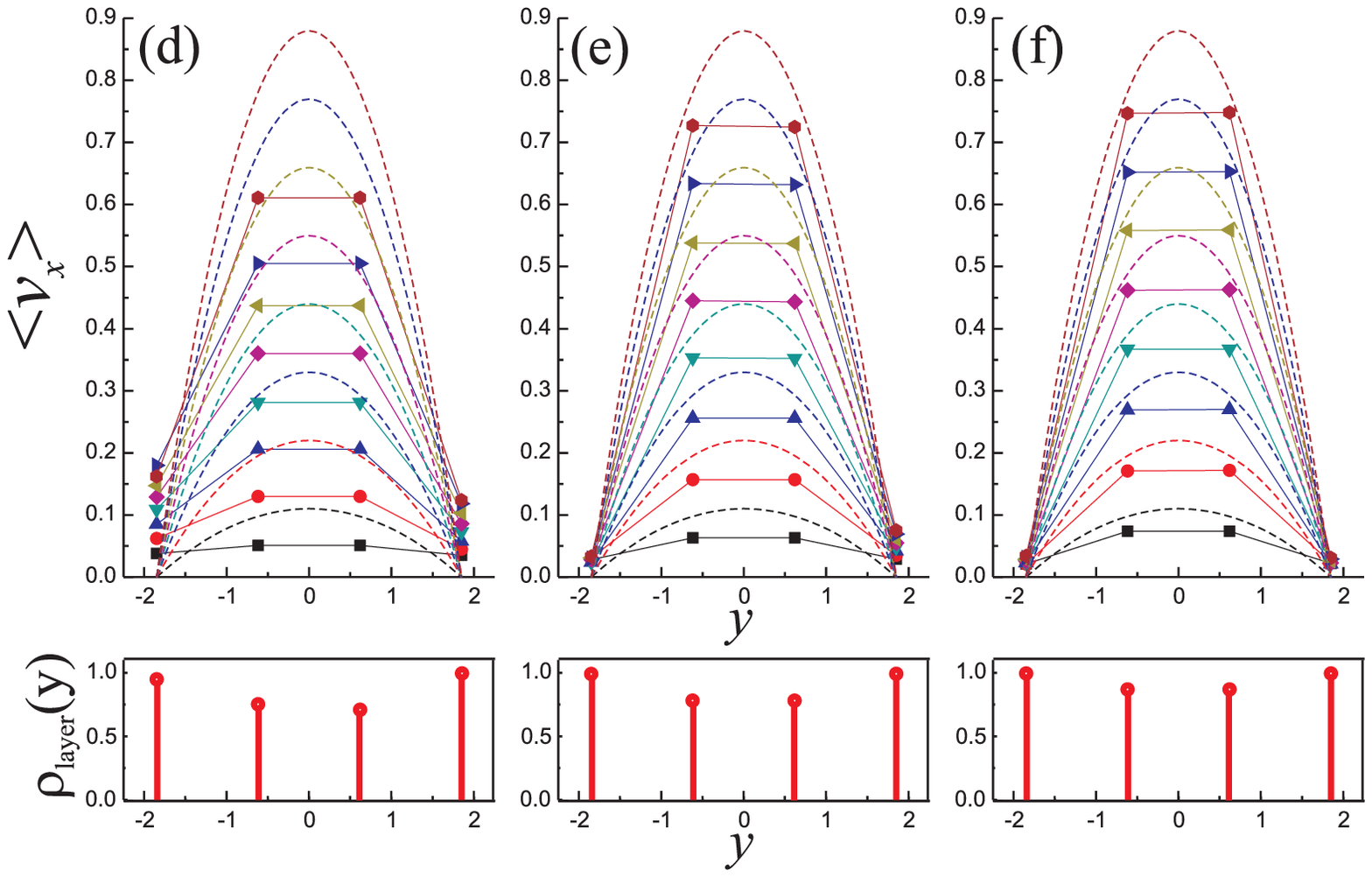}
\end{center}
\vspace{-0.5cm}
\caption{
(Color online)
Top panels:
The profile of average chain velocities $\langle\upsilon_x\rangle$
as a function of the transverse coordinate $y$ for a parabolic
profile of the applied driving force for various values of $\alpha$
($\alpha = 0.1$ to 0.8)
and for the relative width of the channel $L_y/\sigma_{col}=3.7$.
Bottom panels:
The corresponding distributions of the chain densities 
$\rho_{chain}(y)$
shown for comparison (i.e., normalized on $\rho_{max}$, where 
$\rho_{max}$ is the maximum density of the chain in the colloidal 
$n_{l}$-chain configuration). 
The results are shown for the total density $\rho$:
$\rho=0.31 \sigma_{col}^{-2}$ (a),
$\rho=0.41 \sigma_{col}^{-2}$ (a),
$\rho=0.51 \sigma_{col}^{-2}$ (c),
$\rho=0.61 \sigma_{col}^{-2}$ (d),
$\rho=0.71 \sigma_{col}^{-2}$ (e), and
$\rho=0.81 \sigma_{col}^{-2}$ (f).
As in Fig.~\ref{FigLinParVy},
symbols connected by solid lines show average velocities of
chains, and dashed lines show velocities of (non-interacting) chains.
}
\label{FigParabDVy}
\end{figure}

\begin{figure}[btp]
\begin{center}
\vspace*{0.5cm}
\includegraphics*[width=8.5cm]{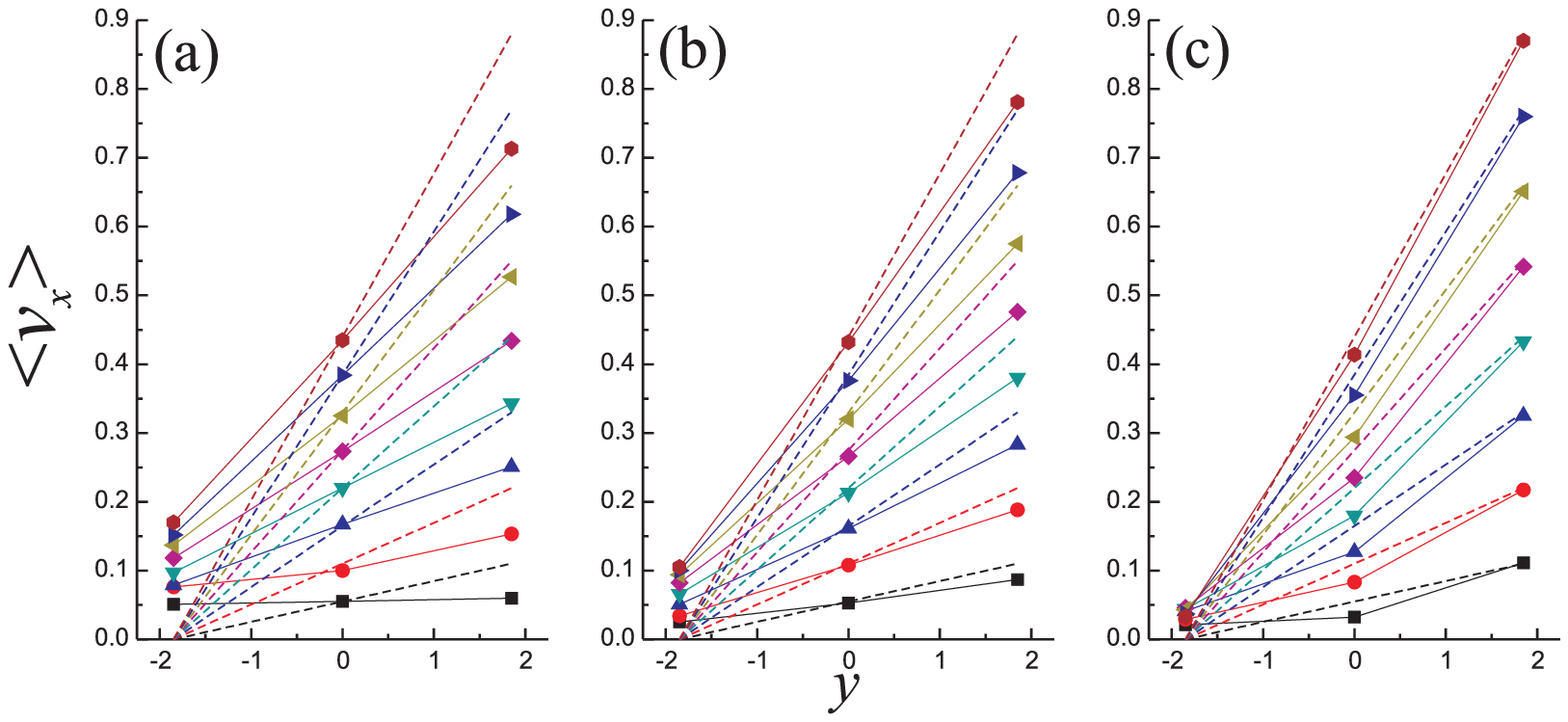}
\includegraphics*[width=8.5cm]{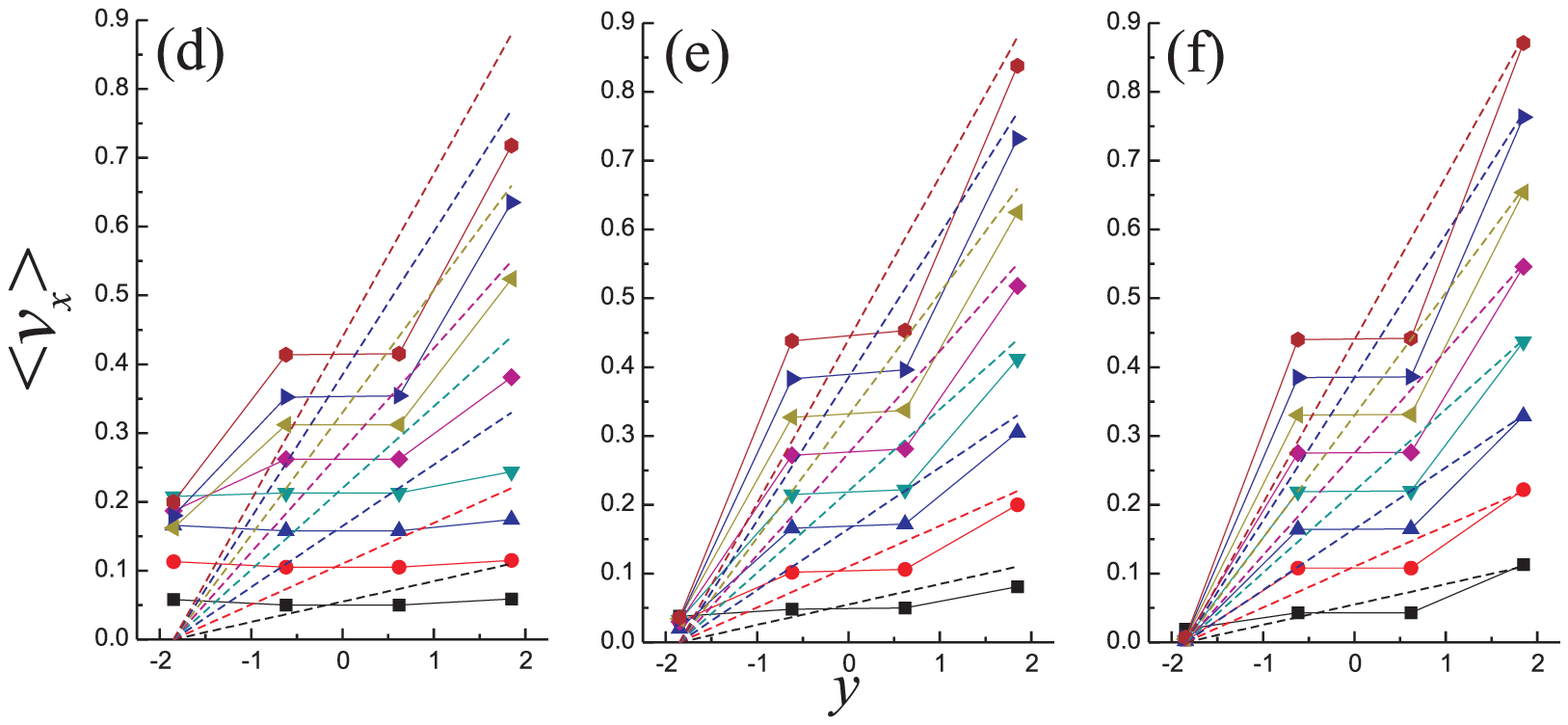}
\end{center}
\vspace{-0.5cm}
\caption{
(Color online)
The profile of average chain velocities $\langle\upsilon_x\rangle$
as a function of the transverse coordinate $y$ for a liner-gradient
profile of the applied driving force for various values of $\alpha$
($\alpha = $0.1 to 0.8)
and for relative width of the channel $L_y/\sigma_{col}=3.7$.
The results are shown for the total density $\rho$:
$\rho=0.31 \sigma_{col}^{-2}$ (a),
$\rho=0.41 \sigma_{col}^{-2}$ (a),
$\rho=0.51 \sigma_{col}^{-2}$ (c),
$\rho=0.61 \sigma_{col}^{-2}$ (d),
$\rho=0.71 \sigma_{col}^{-2}$ (e), and
$\rho=0.81 \sigma_{col}^{-2}$ (f).
As in Fig.~\ref{FigLinParVy},
symbols connected by solid lines show average velocities of
chains, and dashed lines show velocities of (non-interacting) chains.
}
\label{FigLinearDVy}
\end{figure}

In order to study the influence of two competing factors,
the structural factor $\rho_{chain}(y)$ (i.e., the transverse
colloidal density) and the distribution of the external force
$F_{dr}(y)$, we performed simulations of colloidal motion for
various values of the total density $\rho$ in the intermediate
range of values of $\alpha$, $0.1 \leq \alpha \leq 0.8$. 
The average stripe velocity $\bar \upsilon_{x}$ as a function of 
the transverse coordinate $y$ at various values of $\alpha$ and 
different total colloidal density $\rho$ is presented 
in Figs.~\ref{FigParabDVy} and ~\ref{FigLinearDVy}, respectively, 
for parabolic and linear-gradient profiles of the external force 
$F_{dr}(y)$. 
The corresponding distributions of the (normalized) cross-section 
chain density $\rho_{chain}(y)$ are shown in the bottom panels 
of Fig.~\ref{FigParabDVy}. 

As seen from Fig.~\ref{FigParabDVy}(a), for the density 
$\rho=0.31 \sigma_{col}^{-2}$
and $\alpha<0.3$ the colloidal structure consists of three stripes
and moves as a whole (RB motion).
If the colloidal distribution is symmetric with respect
to the channel axis $y = 0$ then the critical value $\alpha$
of the RB-to-plastic motion increases up to $0.5$.
Such an increase of the critical value of $\alpha$ is explained by
the simultaneous locking of motion of the central stripe by two 
adjacent stripes. 
The transition from the $RB$ mode to the plastic mode occurs only
for high enough magnitudes of $F_{max}$ when colloids of the central 
stripe are capable to overcome the potential barriers created by peripheral stripes. 
However, in case of an asymmetric (with respect to the channel 
axis $y = 0$) distribution $\rho_{chain}(y)$, the motion of the 
central stripe is assisted by defects. 
The potential-energy profile asymmetry, together with the 
transverse degree of freedom, leads to such type of motion of 
the central stripe when it avoids obstacles and moves along the 
potential-energy minimum lines, i.e., along serpentine-like 
trajectories. 
This kind of motion provides a lower, as compared to the symmetric 
case, critical value of $\alpha$ of the transition from RB to 
plastic mode.

For density $\rho=0.41$, the RB mode is observed only for 
the lowest value $\alpha=0.1$ (Fig.~\ref{FigParabDVy}(b)). 
For larger densities the RB mode of motion is not observed 
in the range of $\alpha = $0.1 to 0.8 (Fig.~\ref{FigParabDVy}(c)-(f)) 
but it can be realized for even smaller values of $\alpha$. 
Note, however, that for $\rho=0.61 \sigma_{col}^{-2}$ 
(Fig.~\ref{FigParabDVy}(d)) 
(the least dense state of a four-chain phase) at $\alpha=0.1$ 
the motion almost corresponds to the RB mode. 

%
%

Comparison of the cases of a parabolic (Fig.~\ref{FigParabDVy})
and a linear (Fig.~\ref{FigLinearDVy}) driving forces clearly shows 
the difference in the conditions of the manifestation of the RB mode. 
In particular, for a linear-gradient force $F_{dr}(y)$ this mode 
is observed for values $\alpha\leq0.1$ and total density 
$\rho=0.31 \sigma_{col}^{-2}$ 
(i.e., the lowest-$\rho$ three-chain phase), that means a lower
threshold magnitude $F_{max}$ of the transition from elastic to 
plastic mode of colloidal motion (cp. Figs.~\ref{FigParabDVy}(a) 
and~\ref{FigLinearDVy}(a)). 
At the same time, for the lowest-$\rho$ four-chain phase, 
this threshold value is higher $(\alpha\leq0.4)$ in case of 
a linear driving than for a parabolic, as can be seen from 
Figs.~\ref{FigParabDVy}(d) and ~\ref{FigLinearDVy}(d). 
This means that the RB mode manifests itself for: 
(i) a parabolic driving force $F_{dr}(y)$ in the case of 
the lowest-$\rho$ three-chain phase and 
(ii) a linear $F_{dr}(y)$ in case of the lowest-$\rho$ 
four-chain phase. 

The observed behavior is explained as follows.
In case of a parabolic driving of the lowest-$\rho$ three-chain
structure, the driving is applied to the central stripe.
The motion of this stripe is locked by the potential created
by two adjacent stripes.
The most stable is a symmetric configuration, i.e., when two
peripheral stripes are mirror-symmetric with respect to the
axis of the channel $y = 0$, and thus the potential created
by them is double the potential created by a single stripe.
Thus, the symmetric configuration favors the RB mode.
For asymmetric configurations, the RB-to-plastic mode
threshold is lower.
In this case the motion of the central stripe (i.e., for
driving higher than a critical value) is serpentine-like,
i.e., the central stripe deforms in the $y$-direction
following the minimum potential-energy path.
A typical dispersion of the trajectory of the central
stripe in the transverse direction is $\Delta y \approx 0.1$
(while $\Delta y \approx 10^{-4}$ to $10^{-3}$ for the
symmetric case).
As a result, the lowest-$\rho$ three-chain colloidal structure 
is characterized by a relatively high threshold of the RB-to-plastic 
transition, under a parabolic-profile driving 
(see Fig.~\ref{FigParabDVy}(a)). 
In case of a linear-gradient driving of the lowest-$\rho$
three-chain phase (Fig.~\ref{FigLinearDVy}(a)), the maximum
driving force is applied to one of the peripheral chains.
In the RB mode, its motion is locked only by the adjacent
central chain.
When unlocked, the peripheral chain also provides an additional
driving applied to the central chain.
The central chain adjusts itself to follow the minimum
potential-energy path between the two peripheral chains.
This motion is characterized by a rather large transverse
dispersion, $\Delta y \approx 0.2$ to $0.3$, that facilitates
an easy sliding of the three chains with respect to each other.

The situation is quite different in case of
the lowest-$\rho$ four-layer phase shown in
Figs.~\ref{FigParabDVy}(d) and ~\ref{FigLinearDVy}(d).
In the ground-state low-$\rho$ four-chain configurations,
two central chains have lower density than the peripheral
chains (see bottom panels in Figs.~\ref{FigLinearDVy}(d), (e),
and (f)).
Thus, the potential profile created by the central chains
is deeper, and they survive a rather strong shear stress
(similar to the case of low-density 1D chains in a harmonic
potential in the Frenkel-Kontorova model \cite{BraunKivshar}).
Such shear stress is negligible for a parabolic driving
(it appears as a result of transverse instabilities).
Thus the central part of the four-chain structure 
remain in the RB mode for very large values of driving, 
and it also does in case of a linear-gradient driving
(see Figs.~\ref{FigParabDVy}(d), (e), and (f)).
However, the higher-$\rho$ peripheral chains produce
shallower potential-energy profiles, and thus the
friction between them and the central chains is weak
\cite{BraunKivshar}. 
It is clear than that in case of a parabolic driving
the peripheral stripes can easily slide with respect to
the central stripes, and thus the RB-to-plastic mode
threshold is rather low (Fig.~\ref{FigParabDVy}(d)).
In contrast, a relatively weak (i.e., for the same
$F_{max}$) gradient of the driving force in case of
a linear-gradient driving provides a high threshold
of the RB-to-plastic mode transition
(Fig.~\ref{FigLinearDVy}(d)).

\begin{figure}[btp]
\begin{center}
\hspace*{-0.5cm}
\includegraphics*[width=8.5cm]{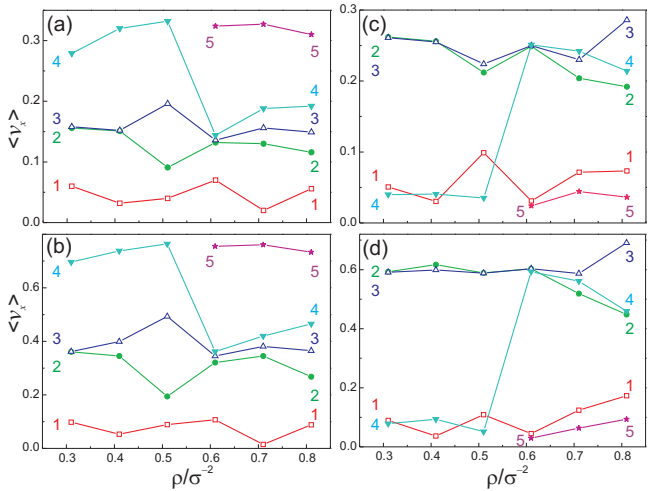}
\end{center}
\vspace{-0.5cm}
\caption{
(Color online)
Average stripe velocities $\langle\upsilon_x\rangle$ 
as a function of the total colloidal density $\rho$ 
for a linear ((a), (b)) and a parabolic ((c), (d)) 
profiles of the applied driving force, for the channel 
width $L_y/\sigma_{col}=5$, and for different driving: 
$\alpha = 0.3$ ((a), (c)) and 
$\alpha = 0.7$ ((b), (d)). 
Different symbols connected by solid lines show 
$\langle\upsilon_x\rangle$ of different chains marked 
for clarity by numbers 1, 2, 3, 4, and 5 
(the numbers are shown by the same color as 
the corresponding curves). 
}
\label{FigVro}
\end{figure}

As we discussed above, the colloidal distribution in a channel
depends on the width of the channel and the total colloidal density.
For certain ``matching'' sets of these parameters,
colloids form well-defined stripe structures
characterized by a maximum value of the order parameter
$\Psi_{n_{l}} = 1$, while small values of the order
parameter $\Psi_{n_{l}} \ll 1$ correspond to non-chained
structure.
It is interesting then to follow the evolution of stripe
velocities when changing the total colloidal density
in the channel \cite{stripes}.
The results of calculations are presented in Fig.~\ref{FigVro}
for a linear-gradient ((a), (b)) and a parabolic ((c), (d))
profile of the driving force and different magnitudes of driving. 
Let us consider first the case of a linear driving.
At low density, $\rho = 0.3 \sigma_{col}^{-2}$, 
the colloids form four stripes 
(marked by numbers 1, 2, 3, and 4 of the same color as the 
corresponding lines/symbols), 
and for all values of $\alpha$ 
(i.e., $\alpha = 0.3$ (a) and 0.7 (b))
shown in Fig.~\ref{FigVro}, two central chains (2 and 3) turn 
out to be locked while the peripheral chains (1 and 4)
(which have a higher stripe density $\rho_{chain}$ and are  incommensurate with the central chains) 
slide with respect to the central chains. 
With increasing $\rho$, 
e.g., $\rho = 0.4 \sigma_{col}^{-2}$ for $\alpha = 0.7$ 
(see Fig.~\ref{FigVro}(b))
the central chains 2 and 3 unlock.
This is explained by decreasing of the inter-colloid distance 
in stripes and thus by shallowing of the potential profile created 
by them. 
Note that for a weaker driving $\alpha = 0.3$ (a), the central 
stripes remain locked at $\rho = 0.4 \sigma_{col}^{-2}$ and 
they unlock at a higher value of $\rho$: 
$\rho = 0.5 \sigma_{col}^{-2}$. 
For even higher density, $\rho = 0.6 \sigma_{col}^{-2}$, 
the inter-shell defects lock the motion of the central chains 
2 and 3 (and partially of all the chains), and the whole 
stripe structure collapses.
If we still increase the colloid density, a new stripe (5) 
originates from the ``disordered'' phase, and for larger 
$\rho$ (i.e., $\rho = 0.8 \sigma_{col}^{-2}$) the motion 
of all five stripes unlock, and they move with individual 
velocities.
Note that the transition of the system from the state with 
$n_{l}$ stripes to the state with $n_{l}+1$ stripes occurs 
always through the disordered phase 
(i.e., when all or some of the stripes lose their identity, 
and colloids of these ``stripes'' move with the same velocity 
as, e.g., stripes 2, 3, and 4 in Fig.~\ref{FigVro}(a) and (b) 
at $\rho = 0.6 \sigma_{col}^{-2}$). 
Thus the disordered phase of the motion serves as a 
``bifurcation point'' that gives rise to a new state with 
$n_{l}+1$ velocity branches. 
The evolution of stripe velocities for a parabolic driving
occurs in a similar way (Fig.~\ref{FigVro}(c) and (d)),
although the degenerate velocities of the central stripes, 
2 and 3 
(and those of the peripheral stripes, 1 and 4, which are 
separated from the velocities of the central stripes by 
a wide gap), unlock only for rather high density. 

It is also interesting to note an oscillating behavior
of the velocity of motion of the peripheral stripes driven
only by the interaction with adjacent stripes.
The oscillation in velocity reflects the oscillation of
the friction force between the stripes with increasing
total colloidal density.

\section{Mobility of stripes and dynamical phases of their motion}

As shown above, the response of the system of colloidal stripes
(i.e., the profile of individual velocities of different stripes)
to the external non-uniform driving is, in general, nonlinear:
the velocity profile of the stripes is not scalable with the
driving (i.e., for all values of the driving, although there
are regions of scalability which will be discussed below).

\begin{figure}[btp]
\begin{center}
\hspace*{-0.5cm}
\includegraphics*[width=8.5cm]{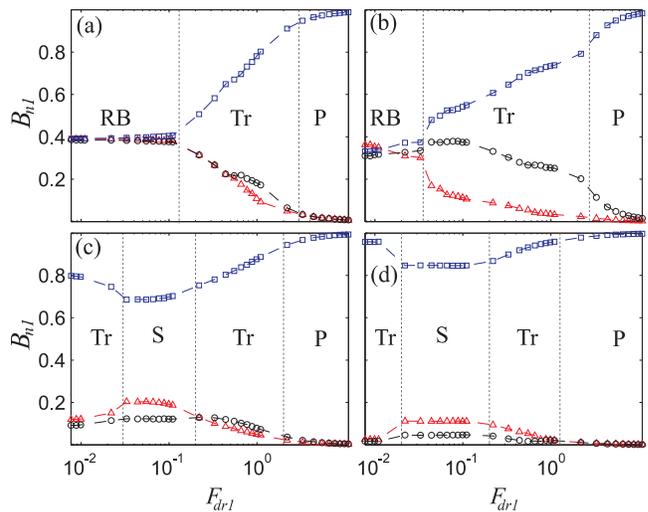} 
\end{center}
\vspace{-0.5cm}
\caption{
(Color online)
The mobility $B_{k,k}$ and mobile response $B_{n,k}(n\neq k)$
of different stripes
as a function of driving force applied to $k$th stripe
($k = 1$)
at various values of the total density:
$\rho=0.45 \sigma_{col}^{-2}$ (a),
$\rho=0.55 \sigma_{col}^{-2}$ (b),
$\rho=0.65 \sigma_{col}^{-2}$ (c), and
$\rho=0.75 \sigma_{col}^{-2}$ (d).
The mobility (mobile response) of different stripes
are shown by different symbols:
$B_{1,1}$ ($\Box$),
$B_{1,2}$ ($\bigcirc$), and
$B_{1,3}$ ($\triangle$).
}
\label{FigB}
\end{figure}

In order to systematically examine the response of the colloidal
system to an external driving, we analyze the mobility of different
chains under the action of a delta-like driving force applied
to {\it one} of the chains:

\begin{equation}
 F_{dr}(y)=F_{drk}(y)=F_{max}\delta(y-y_{k,i}), 
\label{L8}
\end{equation}
where
$\delta(y-y_{k,i}) = 1$, if $y = y_{k,i}$, and
$\delta(y-y_{k,i}) = 0$, if $y \neq y_{k,i}$, 
$y_{k,i}$ is the coordinate of $i$th particle of $k$th stripe 
\cite{driving} 
and $F_{max}$ is the driving force as defined in Eq.~(\ref{L7}). 
We simulated the colloidal motion in a channel of width
$L_y/\sigma_{col}=4$ for various colloidal densities $\rho$. 
For such a channel and considered densities $\rho$, 
the ground state of the system corresponds to a three-stripe 
structure characterized by a high value of the order parameter 
$\Psi_3 \approx 1$. 
Fig.~\ref{FigB} shows the results of calculation for the mobilities versus driving force $F_{drk}$ for different stripes 
$B_{n,k}=\bar{\upsilon}_{xn}/F_{drk}$, 
where $\bar{\upsilon}_{xn}$ is the average velocity of $n$th stripe, 
and $F_{drk}$ is the force applied to $k$th stripe. 
(Here we show only the case when driving is applied to one
peripheral stripe. 
Applying driving to other stripes results in a similar behavior.) 
Note that the mobility is defined by 
$B_{k,k}=\bar{\upsilon}_{xk}/F_{drk}$, 
while $B_{n,k}$ for $k \neq n$ is a mobile response of the stripes. 

As seen from Fig.~\ref{FigB}, the mobility as a function of 
the applied force exhibits a variety of modes of motion. 
First, we stress the most general features of the behavior. 
A stripe mobility $B_{k,k}$ tends to unity in the limit of 
large applied force $F_{drk}$, and mobile responses $B_{n,k}$, 
correspondingly, vanish, as seen from Fig.~\ref{FigB}. 
This limit corresponds to the plastic mode of motion. 
Another general conclusion refers to the RB-to-plastic 
transition: this transition shifts towards lower drivings 
with increasing the total colloidal density 
(cp. Figs.~\ref{FigB}(a) and (b)). 
As we discussed above, a growth of the colloidal density 
results in a hardening of the rigidity of the elastic stripes 
\cite{BraunKivshar} and a simultaneous weakening of the 
inter-stripe friction. 

However, apart from these general features, the mobilities display
other important peculiarities.
To discuss these, let us define dynamical phases corresponding
to different regimes of colloidal motion in channels.
We denote the first regime as ``RB'' (``rigid body'', or elastic
motion).
This regime is most pronounced for low colloidal density,
$\rho=0.45 \sigma_{col}^{-2}$ (Fig.~\ref{FigB}(a)), where
it extends to $F_{dr1} \approx 1.05$.
For
$\rho=0.55 \sigma_{col}^{-2}$ (Fig.~\ref{FigB}(b)) the threshold
value reduced by a factor of five, $F_{dr1} \approx 0.2$.
For higher densities, the RB-to-plastic mode transition
shifts towards very low values of $F_{dr1}$.
The RB regime is followed by a broad transition region (``Tr'')
characterized by a monotonic increase of the mobility $B_{k,k}$
(and monotonic decrease of the mobile responses $B_{n,k}$).
While this growth is close to linear in Fig.~\ref{FigB}(a), for
higher density it displays oscillations in Fig.~\ref{FigB}(b).
Remarkably, for higher colloidal densities these oscillations
turn to regions of {\it decreasing} colloid mobility $B_{k,k}$
versus driving force.
Note that this regime (denoted as ``S'') is similar to the
dynamically-ordered phase of vortex motion in superconductors 
with arrays of regular pinning sites \cite{dynphfn,dynphwe}.
This ordered phase follows a disordered phase with a higher
density of average vortex flow and thus leads to the appearance
of a negative differential resistivity (NDR) part of N-type
in the VI-curve \cite{dynphwe}.
In our case of colloidal motion in a narrow channel,
we observe a similar effect characterized by a slowing down
of the motion of the stripe driven by an external force
(with a simultaneous increase of the velocity of the
adjacent stripes) when increasing the applied driving force.
In this region, the system of colloids displays a partial
``reentrant'' behavior 
(cp. Refs.~\cite{dzubiellajpcm,dzubiellapre}) 
when, being melted by increasing shear 
stress, it starts evolving towards a solid phase with further
increasing driving.
Note that this dynamically-induced ``solidification''
has common features with the recently discovered transition
``freezing by heating'' \cite{fbhprl,fbhnat} where by increasing
temperature it leads to the crystallization of moving repulsive
interacting particles which are in a molten state.

The mechanism of the observed colloid solidification can be
understood as follows.
As we discussed above, the longitudinal motion of colloidal
stripes is accompanied by transverse oscillations of the
stripes' trajectories (i.e., serpentine-like motion)
related to the asymmetry of the potential-energy profiles
created by the adjacent stripes.
For low drivings (but larger than the critical driving force 
of the RB-to-plastic transition), the stripe driven by 
an external force moves along some serpentine-like 
trajectory. 
When moving, the stripe itself deforms adjusting to the
potential-energy landscape.
In its turn, it elastically deforms other stripes due to
the interaction with colloids in the adjacent stripes.
At low velocities, colloids in adjacent stripes relax
to the initial state thus providing low friction between
the stripes.
Increasing the velocity of motion of the driven stripe (at high
colloid density) leads to increasing rate of the inter-stripe
collisions and to the development of instabilities in the
transverse direction.
This results in an increase of the dynamical friction between
stripes which explains the observed decrease of the mobility
$B_{k,k}$ (and increase of the mobile response $B_{n,k}$),
i.e., the appearance of ``S'' phase.
Very large driving, however, straightens the trajectories of
motion of the driven stripe.
In the limit of $F_{dr} \to \infty$, the driven stripe moves
along unrelaxed in the $y$-direction straight trajectory, 
and the mobility $B_{k,k} \to 1$ (while the mobile response
$B_{n,k} \to 0$), as shown in Figs.~\ref{FigB}(c), (d).

\section{Conclusions}

We have studied the dynamics of colloids driven by an external 
non-uniform force in a narrow channel. 
We focused on colloidal densities which provide well-defined 
colloidal stripe structures. 
In the limit of a very low driving force the system moves as 
a rigid body (elastic motion) independent of the profile of the 
driving force. 
In the opposite limit of a very strong driving force, the profile 
of the average velocities of colloidal stripes follow the profile 
of the applied driving (plastic regime). 
While these limiting-case results are easily understandable, 
the dynamics of colloidal stripes in the general case is rather 
complex. 
It is governed by the interplay of several factors, i.e.: 
(i) the magnitude and profile of driving force; 
(ii) the total density of the colloidal system and 
the distribution of the colloidal density in stripes, 
and, as a consequence, 
(iii) commensurability effects between the adjacent chains. 
We have shown that depending on the density and number of 
stripes, the transition from elastic to plastic motion 
occurs at different values of the driving force. 
For example, the lowest-density three-stripe colloidal 
configuration is shown to be more robust with respect to 
a parabolic driving than to a linear-gradient driving, 
while for the lowest-density four-stripe configuration
the situation is opposite. 
This result is explained by the colloidal density 
distribution over the central (lower density) and the peripheral 
(higher density) stripes and commensurability effects. 
We have analyzed the mobility of colloidal stripes 
and have identified the dynamical phases of their motion. 
In particular, it has been shown that the transition 
from elastic (rigid-body-motion phase) to plastic mode, 
depending on the density of colloids, could be either 
monotonic (i.e., characterized by a gradual increase of 
the mobility), or it could contain an NDR-type part 
(i.e., characterized by a drop of mobility versus 
driving force). 
This unusual ``solidification'' is related to the 
dynamically-induced increase of the friction in the 
colloidal system and is similar to the recently 
discovered ``freezing by heating'' transition. 
The results of our study, with corresponding changes, 
can also be applied to other systems of interacting 
particles driven by a non-uniform force in narrow channels, 
e.g., in physics or biology.

\section{Acknowledgments}

This work was supported by the ``Odysseus'' program of the 
Flemish Government and Flemish Science Foundation (FWO-Vl). 
V.R.M. acknowledges the support by the EU Marie Curie 
project, Contract No. MIF1-CT-2006-040816.


\begin{references}
\bibitem{stripes}
Note that although stripes are not well defined for low values of the
order parameter $\Psi_{n_{l}} \ll 1$, we still call them ``stripes''.
We start from well-defined (i.e., narrow) chains and then follow the
evolution of the stripes even though they broaden.
\bibitem{driving} 
This means that the driving force is applied to all the particles of 
$k$th stripe, even if their individual transverse coordinates $y_{k,i}$ 
deviate from the average $y$-coordinate of the stripe.
\end{references}
\end{document}